\documentclass{PoS}

\title{General properties of logarithmically divergent one-loop lattice
Feynman integrals}

\ShortTitle{Logarithmically divergent one-loop lattice integrals}

\author{\speaker{Jongjeong Kim}\\
        Department of Physics and Astronomy, Seoul National University,
        Seoul, 151-747, South Korea\\
        E-mail: \email{rvanguard@phya.snu.ac.kr}}

\author{David H. Adams\\
        Department of Physics and Astronomy, Seoul National University,
        Seoul, 151-747, South Korea\\
        E-mail: \email{dadams@phya.snu.ac.kr}}

\author{Weonjong Lee\\
        Frontier Physics Research Division and Center for Theoretical Physics,\\   
        Department of Physics and Astronomy, Seoul National University,
        Seoul, 151-747, South Korea\\
        E-mail: \email{wlee@phya.snu.ac.kr}}

\abstract{We prove that logarithmically divergent one-loop lattice Feynman 
integrals have the
general form $I(p,a)=f(p)\log(aM)+g(p,M)$ up to terms which vanish for lattice
spacing $a\to0$. Here $p$ denotes collectively the external momenta and $M$ is an 
arbitrary mass scale. The $f(p)$ is shown to be universal and to coincide with the
analogous quantity in the corresponding continuum integral (regularized, e.g.,
by momentum cut-off). This is essential for
universality of the lattice QCD beta-function and anomalous dimensions of 
renormalized lattice operators at one loop. The result and argument presented
here are simplified versions of ones given in arXiv:0709.0781. A noteworthy 
feature of the argument here is that it
does not involve Taylor expansion in external momenta, hence infra-red 
divergences associated with that expansion do not arise.}

\FullConference{The XXV International Symposium on Lattice Field Theory\\
		 July 30-4 August 2007\\
		 Regensburg, Germany}

\def\be{\begin{eqnarray}}
\def\ee{\end{eqnarray}}

\def\hk{\hat{k}}
\def\hd{\hat{d}}
\def\ha{\hat{a}}

\begin{document}

\section{Introduction}

Lattice folklore asserts that the one-loop beta-function and anomalous 
dimensions of one-loop renormalized operators are universal in lattice formulations 
of QCD and coincide with the corresponding continuum quantities.
While this is certainly the case for the explicit lattice formulations 
considered to date, no general proof for general lattice QCD 
formulations has been given previously. The renormalization factors that one needs 
to consider in connection with this
are given by logarithmically divergent one-loop lattice Feynman
integrals $I(p,a)$. In this paper we derive a general structural result for
such integrals which provides a basis for proving the mentioned folklore assertions.
We show that these integrals have the general structure
\be
I(p,a)=f(p)\log(aM)+g(p,M)
\label{1}
\ee
up to terms which vanish for lattice spacing $a\to0$. Here $p$ denotes collectively
the external momenta and $M$ is a mass scale which may be chosen arbitrarily. 
The key features are (i) the lattice spacing dependence enters exclusively through
the $\log(aM)$ term, and (ii) the factor $f(p)$ can be expressed as a convergent 
continuum integral and is therefore universal; moreover,
\be
f(p)=f^{(c)}(p)
\label{2}
\ee
where $f^{(c)}(p)$ is the analogous factor in the 
corresponding continuum integral, which has the structure
\be
I^{(c)}(p,\Lambda)=f^{(c)}(p)\log(M/\Lambda)+g^{(c)}(p,M)
\label{3}
\ee
when regularized by a momentum cut-off $\Lambda$. (Analogous statements hold when
dimensional regularization is used; see Ref.\cite{log}.)

A stronger version of the structural results (\ref{1})--(\ref{2}) was recently 
proved in Ref.\cite{log}.\footnote{This structure was expected 
\cite{Symanzik,Luscher-Weisz(pert)} but we emphasize that no explicit general
proof had previously been given.}\footnote{Some of the techniques and results of
Ref.\cite{log} were developed earlier in a special case in Ref.\cite{prev}}
The argument there gave more information on the structure
of $g(p,M)$ in (\ref{1}). However, (\ref{1})--(\ref{2}) alone can be established by
a simpler argument than the one given there, which we will present here. 
In particular, the argument avoids the usual procedure of separating off
the leading term in the expansion of $I(p,a)$ in $p$ and having to deal with its
infra-red divergence.

The argument proceeds in two main steps: First we define
\be
f(p):=\lim_{a\to0}\ a\frac{d}{da}I(p,a)
\label{4}
\ee
and show that it is finite and given by a convergent continuum integral coinciding 
with $f^{(c)}(p)$. Next we define
\be
g(p,M):=\lim_{a\to0}\ \Big(I(p,a)-f(p)\log(aM)\Big)
\label{5}
\ee
and prove that it is finite.
Together these imply the structural results (\ref{1})--(\ref{2}).

\section{The setup}

One-loop lattice Feynman integrals have the following general form:
\be
I(p,a)&=&\int_{-\pi/a}^{\pi/a}d^4k\,\frac{V(k,p,a)}{C(k,p,a)}
\label{2.1} \\
V(k,p,a)=\frac{1}{a^m}F(ak,ap)&&,\quad\quad C(k,p,a)=\frac{1}{a^n}G(ak,ap)
\label{2.2}
\ee
The lattice degree \cite{R} of $V(k,p,a)$ in $k$ can be characterized as follows.
Let $r$ denote the order of the first non-vanishing term in the $t$-expansion
of $F(k,tp)$ around $t=0$, 
\be
F(k,tp)=t^rF_0(k,p)+t^{r+1}F_1(k,p,t)
\label{2.3}
\ee
Then $V(\lambda k,p,\frac{a}{\lambda})=(\frac{\lambda}{a})^mF(ak,\frac{a}{\lambda}p)
\sim\lambda^{m-r}$ for $\lambda\to\infty$, hence the lattice degree of $V$ is
$\hd_V=m-r$ \cite{R}.

We assume that $C(k,p,a)$ arises as a product of lattice propagators as described
in Reisz's work \cite{R}. Then the expansion of $G(k,tp)$ in $t$ has non-vanishing 
zero-order term:
\be
G(k,tp)=G(k,0)+tG_1(k,p,t)
\label{2.4}
\ee
hence
$C(\lambda k,p,\frac{a}{\lambda})=(\frac{\lambda}{a})^nG(ak,\frac{a}{\lambda}p)
\sim\lambda^n$ for $\lambda\to\infty$, so the lattice degree of $C$
is $\hd_C=n$.
The divergence degree of the lattice integral is then given by \cite{R}:
\be
\hd_I=4+\hd_V-\hd_C=4+m-r-n\,.
\label{2.5}
\ee
$V(k,p,a)$ and $C(k,p,a)$ are assumed to have finite continuum limits:
\be
V(k,p,a)\stackrel{a\to0}{\longrightarrow}P(k,p)\quad,\quad
C(k,p,a)\stackrel{a\to0}{\longrightarrow}E(k,p)
\label{2.6}
\ee
Then, in light of (\ref{2.2}), it follows that $P(k,p)$ and $E(k,p)$ are
homogeneous polynomials in $(k,p)$ of degrees $m$ and $n$, respectively.
We denote the usual degree of $P(k,p)$ in $k$ by $d_P$, with $d_E$ defined
analogously for $E(k,p)$. Note that $d_P\le\hd_V$ while $d_E=n=\hd_C$ \cite{R}.

We henceforth specialize to the logarithmically divergent case, i.e., $\hd_I=0$,
and proceed to carry out the two steps discussed in the introduction for 
establishing the claimed structural results for $I(p,a)$.

\section{Step 1}

Changing variables to $\hk=ak$ in (\ref{2.1}) and using $\hd_I=4+m-r-n=0$ we find
\be
I(p,a)=\int_{-\pi}^{\pi}d^4\hk\,\frac{a^{-r}F(\hk,ap)}{G(\hk,ap)}
\label{3.1}
\ee
and consequently
\be
a\frac{d}{da}I(p,a)
&=&\int_{-\pi}^{\pi}d^4\hk\,a\frac{d}{da}\bigg(\frac{a^{-r}F(\hk,ap)}{G(\hk,ap)}
\bigg) \nonumber \\
&=&\int_{-\pi}^{\pi}d^4\hk\,\frac{d}{dt}\bigg(\frac{(ta)^{-r}F(\hk,tap)}
{G(\hk,tap)}\bigg)_{t=1} \nonumber \\
&=&\int_{-\pi/a}^{\pi/a}d^4k\,\frac{d}{dt}\bigg(\frac{t^{-r}V(k,tp,a)}
{C(k,tp,a)}\bigg)_{t=1} 
\label{3.2}
\ee
We are going to show that the last integral has lattice divergence degree 
$\le\hd_I-1=-1$, which will allow us to apply to it the lattice power-counting 
theorem of Ref.\cite{R}. This is an immediate consequence of the following:

\vspace{1ex}

\noindent (A) $\frac{d}{dt}(t^{-r}V(k,tp,a))$ has divergence degree 
$\le\hd_V-1$.

\vspace{1ex}

\noindent (B) $\frac{d}{dt}C(k,tp,a)$ has divergence degree $\le d_C-1$

\vspace{1ex}

\noindent {\em Proof of (A)}: Using (\ref{2.2})--(\ref{2.3}) we have
\be
t^{-r}V(\lambda k,tp,\frac{a}{\lambda})
&=&\Big(\frac{\lambda}{a}\Big)^mt^{-r}F(ak,\frac{t}{\lambda}ap)
\nonumber \\
&=&\Big(\frac{\lambda}{a}\Big)^mt^{-r}\Big(
\Big(\frac{t}{\lambda}\Big)^rF_0(ak,ap)+\Big(\frac{t}{\lambda}\Big)^{r+1}
F_1(ak,ap,\frac{t}{\lambda})\Big)
\label{3.3}
\ee
The $t$-dependence cancels out in the first term so it 
vanishes under $\frac{d}{dt}$. Hence the $\lambda\to\infty$ behavior is determined
by the second term, which is $\sim\lambda^{m-r-1}$ or slower. 

\vspace{1ex}

\noindent {\em Proof of (B)}: 
This follows by a similar argument, using (\ref{2.4}) to write
\be
C(\lambda k,tp,\frac{a}{\lambda})=\Big(\frac{\lambda}{a}\Big)^n
\Big(G(ak,0)+\frac{t}{\lambda}G_1(ak,ap,\frac{t}{\lambda})\Big) 
\label{3.4}
\ee
and noting that the 
leading term is again $t$-independent and hence vanishes under $\frac{d}{dt}$.

\vspace{1ex}

Having established that the lattice integral expression (\ref{3.2})
for $a\frac{d}{da}I(p,a)$ has lattice divergence degree $\le -1$ we can invoke
Reisz's lattice power-counting theorem \cite{R} to conclude that its $a\to0$
limit is finite and given by the corresponding continuum integral:
\be
f(p):=\lim_{a\to0}\ a\frac{d}{da}I(p,a)
=\int_{-\infty}^{\infty}d^4k\,\frac{d}{dt}\bigg(\frac{t^{-r}P(k,tp)}{E(k,tp)}
\bigg)_{t=1}
\label{3.5}
\ee
Thus $f(p)$ is given by a convergent continuum integral and hence is universal as 
claimed.\footnote{A simpler continuum integral expression for $f(p)$ is obtained
in Eq.(4.7) of Ref.\cite{log}. It would be better to use that one rather than
the present one for evaluating $f(p)$ in practice.} 
It remains to show $f(p)=f^{(c)}(p)$. 
From (\ref{3}) we have
\be
f^{(c)}(p)=\lim_{\Lambda\to\infty}\ -\Lambda\frac{d}{d\Lambda}I^{(c)}(p,\Lambda)
\label{3.6}
\ee
Changing integration variable to $\hk=k/\Lambda$ in the continuum
integral we find
\be
I^{(c)}(p,\Lambda)=\int_{-\Lambda}^{\Lambda}d^4k\,\frac{P(k,p)}{E(k,p)}
=\int_{-1}^{1}d^4\hk\,\frac{\Lambda^rP(\hk,p/\Lambda)}{E(\hk,p/\Lambda)}
\label{3.7}
\ee
where we have used that the previously mentioned homogeneity of $P(k,p)$ and 
$E(k,p)$ in $(k,p)$ implies $P(\Lambda\hk,p)=\Lambda^mP(\hk,p/\Lambda)$ and 
$E(\Lambda\hk,p)=\Lambda^nE(\hk,p/\Lambda)$, and also used 
$0=\hd_I=4+m-r-n$. It follows that
\be
-\Lambda\frac{d}{d\Lambda}I^{(c)}(p,\Lambda)
&=&\int_{-1}^{1}d^4\hk\,\frac{d}{dt}\bigg(\frac{(\Lambda/t)^rP(\hk,pt/\Lambda)}
{E(\hk,pt/\Lambda)}\bigg)_{t=1} \nonumber \\
&=&\int_{-\Lambda}^{\Lambda}d^4k\,\frac{d}{dt}\bigg(
\frac{t^{-r}P(k,tp)}{E(k,tp)}\bigg)_{t=1}
\label{3.8}
\ee
This reduces in the $\Lambda\to\infty$ limit to our previous expression
(\ref{3.5}) for $f(p)$, thereby showing that $f(p)=f^{(c)}(p)$ as claimed.

\section {Step 2}

Our goal now is prove that the limit (\ref{5}) is finite. To this end we set
\be
\ha=aM
\label{4.1}
\ee
and define
\be
f(p,\ha):=a\frac{d}{da}I(p,a)=\ha\frac{d}{d\ha}I(p,\ha/M)
\label{4.2}
\ee
which has $f(p)$ as its $\ha\to0$ limit, cf. \S3. Then the quantity in (\ref{5}) 
whose limit we need to consider can be expressed as
\be
I(p,a)-f(p)\log(aM)&=&I(p,\ha/M)-f(p)\log(\ha) \nonumber \\
&=&-\int_{\ha}^1db\,\frac{1}{b}\Big(f(p,b)-f(p)\Big)+I(p,1/M)
\label{4.3}
\ee
To show that this has finite $a\to0$ limit, or equivalently, finite $\ha\to0$ limit,
we need to show that the integral on the right-hand side remains finite for
$\ha\to0$. For this we need information on how quickly $f(p,b)$ approaches its 
continuum limit $f(p)$ for $b\to0$. To obtain this, we note that $f(p,\ha)$ can be 
expressed as a convergent lattice integral: Changing variables to $\hk=k/M$
in (\ref{3.2}) we find, using (\ref{2.2}) and $\hd_I=4+m-r-n=0$,
\be
f(p,\ha)=\int_{-\pi/\ha}^{\pi/\ha}d^4\hk\,\frac{d}{dt}\bigg(
\frac{(t/M)^{-r}V(\hk,tp/M,\ha)}
{C(\hk,tp/M,\ha)}\bigg)_{t=1}
\label{4.4}
\ee
which is seen to have divergence degree $\le-1$ by the same argument as for 
(\ref{3.2}). An extension of the lattice power-counting theorem proved in
Ref.\cite{log} now tells that $f(p,\ha)-f(p)$ vanishes at least as fast as
$\sim\ha\log(1/\ha)$ for $\ha\to0$. Applying this with $\ha$ replaced by $b$
allows us to conclude that the integral
\be
\int_{\ha}^1db\,\frac{1}{b}|f(p,b)-f(p)|
\label{4.5}
\ee
remains finite in the $\ha\to0$ limit.\footnote{Note that $\frac{1}{b}(b\log(1/b))
=-\log(b)=\frac{d}{db}(-b\log(b)+b)$.} By Lebesgue's ``theorem of dominated 
convergence'' the integral continues to have a well-defined finite limit when the 
integrand is replaced by $\frac{1}{b}(f(p,b)-f(p))$. Hence the $a\to0$ limit of 
(\ref{4.3}) is finite, i.e., $g(p,M)$ in (\ref{5}) is finite as claimed. 
This completes the derivation of the structural results (\ref{1})--(\ref{2}).

\section{Concluding remarks}

Besides depending on the external momenta $p$, the lattice integral $I(p,a)$ may 
also depend on masses (e.g., fermion masses) if the Feynman diagram involves
massive propagators. The dependence of $I(p,a)$ on such masses, which we denote
collectively by $m$, is easily described. Their presence in the lattice
integrand can be described by replacing $F(ak,ap)\to F(ak,ap,am)$ and 
$G(ak,ap)\to G(ak,ap,am)$ in (\ref{2.2}), and from this it is clear that the 
$m$-dependence in the structural results (\ref{1})--(\ref{2}) can be indicated by
replacing $p\to(p,m)$ in the expressions there. Or we can simply take $p$ to
denote collectively the external momenta {\em and} masses.

The structural results (\ref{1})--(\ref{2}) immediately imply universality of the 
anomalous dimensions of one-loop renormalized lattice operators in the usual
case where the renormalization factor is given by a logarithmically divergent 
lattice Feynman integral. Universality of the one-loop beta-function in lattice QCD 
requires a bit more elaboration since the individual lattice Feynman diagrams 
relevant for computing the lattice beta-function can have stronger divergences 
(linear and quadratic). 
However, these can be combined into logarithmically divergent integrals
by exploiting lattice BRST symmetry and lattice hypercubic symmetries 
\cite{R(NPB),Kawai}, after which the structural results here can be applied.
Or we can chose to define the renormalized coupling via the 4-gluon vertex;
then, since the one-loop corrections to this vertex in lattice QCD are all 
either logarithmically divergent or finite, our structural results can be 
immediately applied. This will be discussed explicitly in a forthcoming publication.

The general conditions under which the structural results here are derived can be 
summarized by saying that they are the same as the ones in Reisz's work on
the lattice power-counting theorem \cite{R} (but with $\hd_I=0$ rather than
$\hd_I<0$), except that we do not need a certain technical condition on the lattice
propagators that he required.\footnote{The doubler-free condition is required
both in Reisz's work and our treatment. This excludes naive and staggered fermions.
However, it is nevertheless often possible to apply the results here to lattice
integrals involving these. This can be done when symmetries of the integrand 
allow the integral to be rewritten as 
$\int_{-\pi/a}^{\pi/a}d^4k\,(\cdots)=N_t\int_{-\pi/2a}^{\pi/2a}d^4k\,(\cdots)$ 
where $N_t$ is the number of tastes ($=16$ for naive fermions and $4$ for staggered 
fermions). For example, this is possible for the naive or staggered fermion loop 
contribution to the one-loop vacuum polarization \cite{Weisz}.}
This is because it was possible to prove the
(extended) lattice power-counting theorem in the one-loop case in Ref.\cite{log}
without invoking this condition (see \cite{log} for further discussion).
To deal with the possibility of infrared divergences when the lattice propagators 
are massless, Reisz introduced in Ref.\cite{R(massless)} the notion of infrared
lattice divergence degree and proved infrared finiteness when this degree is 
strictly negative. We also require this condition here in the case where the lattice
propagators are massless. In practice, for the present one-loop case, it usually
means that the external momenta $p$ cannot all be zero.

Finally, as a guide for attempting to extend the structural results to 
divergent multi-loop lattice integrals with general divergence degree, we note that 
the general structure is expected to be as follows 
\cite{Symanzik,Luscher-Weisz(pert)}:
\be
I(p,a)=a^{-\omega}\sum_{n=0}^{\infty}\sum_{m=0}^l c_{mn}(p)a^n(\log\,a)^m
\label{5.1}
\ee
Here $\omega$ is the divergence degree of the integral and $l$ is the number of 
loops. In the one-loop case this can be expressed as a generalization of (\ref{1}):
\be
I(p,a)=a^{-\omega}(f(p,a)\log(a)+g(p,a))
\label{6.2}
\ee
where $f(p,a)=f_0(p)+f_1(p)a+f_2(p)a^2+\dots$ and 
$g(p,a)=g_0(p)+g_1(p)a+g_2(p)a^2+\dots$.

\section{Acknowledgements}

This research is supported by the KICOS international cooperative
research program (KICOS grant K20711000014-07A0100-01410), 
by the KRF grant KRF-2006-312-C00497, by the BK21 program of 
Seoul National University, and by the DOE SciDAC-2 program.

\end{document}